%% file: manuscript.tex
\documentclass[journal=jacsat,manuscript=article,onecolumn]{elsarticle}
\usepackage[latin9]{inputenc}
\usepackage{listings}
\usepackage{setspace}
\usepackage{subcaption}
\usepackage{xcolor}
\definecolor{dark-red}{rgb}{0.4,0.15,0.15}
\definecolor{dark-blue}{rgb}{0.15,0.15,0.4}
\definecolor{medium-blue}{rgb}{0,0,0.5}



\title{\emph{ProtoMD}: A Prototyping Toolkit for Multiscale Molecular Dynamics}

\author[2,3]{Endre Somogyi\corref{note}}
\author[1,3]{Andrew Abi Mansour\corref{note}}
\author[1,3]{Peter J. Ortoleva\corref{cor}}

\address[1]{Department of Chemistry, Indiana University, Bloomington}
\address[2]{Department of Physics, Indiana University, Bloomington}
\address[3]{Center for Theoretical and Computational Nanoscience, Indiana University, Bloomington}
\cortext[cor]{Corresponding author: ortoleva@indiana.edu}
\cortext[note]{Both authors contributed equally to this work}

\lstdefinestyle{verbo}{
    basicstyle=\scriptsize\ttfamily,
    breaklines=true,
    breakatwhitespace=true,
    tabsize=1,
    resetmargins=true,
    xleftmargin=0pt,
    frame=none,
    showspaces=false,
    showstringspaces=false,
}

\date{\today}
\begin{document}

\onehalfspacing

\begin{frontmatter}
\begin{abstract}
\emph{ProtoMD} is a toolkit that facilitates the development of algorithms for multiscale molecular dynamics (MD) simulations.
It is designed for multiscale  methods which capture the dynamic transfer of information across multiple spatial scales,
such as the atomic to the mesoscopic scale, via coevolving microscopic and coarse-grained (CG) variables.
\emph{ProtoMD} can be also be used to calibrate parameters needed in traditional CG-MD methods.
The toolkit integrates `GROMACS wrapper' to initiate MD simulations, and `MDAnalysis' to analyze and manipulate trajectory files.
It facilitates experimentation with a spectrum of coarse-grained variables, prototyping rare events (such as chemical reactions),
or simulating nanocharacterization experiments such as terahertz spectroscopy, AFM,
nanopore, and time-of-flight mass spectroscopy. \emph{ProtoMD} is written in python and is freely available under the GNU General
Public License from github.com/CTCNano/proto\_md.
\end{abstract}


\end{frontmatter}

{\bf PROGRAM SUMMARY/NEW VERSION PROGRAM SUMMARY}

\begin{small}
\noindent
{\em Manuscript Title:}   ProtoMD: A Prototyping Toolkit for Multiscale Molecular Dynamics                                    \\
{\em Authors:}    Endre Somogyi, Andrew Abi Mansour, and Peter J. Ortoleva                                            \\
{\em Program Title:}       ProtoMD                                   \\
{\em Journal Reference:}                                      \\
{\em Catalogue identifier:}                                   \\
{\em Licensing provisions:}       GPL v3 (or above)                            \\
{\em Programming language:} python 2.7.3                       \\
{\em Computer:}  x86 / x86\_64                                             \\
{\em Operating system:}   Linux                               \\
{\em RAM:} Depends on the size of the system being simulated and duration of the simulation (few MBs to TBs) \\
{\em Number of processors used:}      12 - 128                        \\
{\em Supplementary material:}                                 
{\em Classification:}                                         \\
  Computational Methods \\
{\em External routines/libraries:}                            \\
  GROMACS, MDAnalysis, GromacsWrapper, numpy, scipy \\
{\em Subprograms used:}                                       \\
{\em Nature of problem:}\\
  Prototyping multiscale coarse-grained algorithms for molecular dynamics
   \\
{\em Solution method:}\\
  Combining the open-source GROMACS molecular dynamics package and the python-based MDAnalysis
  library for running, debugging, and analyzing multiscale simulations \\
{\em Restrictions:}\\
 The system under study must be characterized by a clear separation of timescales;
otherwise, the multiscale algorithm fails to capture the slowly-varying modes.
   \\
{\em Unusual features:}\\
   \\
{\em Additional comments:}\\
   \\
{\em Running time:}\\
  Depending on the problem size, simulations can take few hours to months.
   \\
\end{small}

\doublespacing

\input{Introduction.tex}

\input{Theoretical_Framework.tex}

\input{Design_and_Implementation.tex}

\input{User_Guide.tex}

\input{Acknowledgments.tex}

\bibliographystyle{unsrt}
\bibliography{refs}

\end{document}

%% file: Introduction.tex
\section{Introduction}

Supramolecular assemblies of contemporary interest include viruses,
vaccine nanoparticles, nanocapsules for drug delivery, nanomaterials,
and light-harvesting fibers. These systems have been studied via all-atom \cite{CHARMM1983,GROMACS2005,NAMD2005,LAMMPS1995,AMBER}
and coarse-grained (CG) \cite{Broughton1998,Schulten2006,Schulten2007,Sansom2007} molecular dynamics (MD). MD
software uses physical concepts
and software engineering to make simulations accurate and efficient
via modern hardware (e.g., GPUs and multicore platforms). However, simulating supramillion-atom systems
over long time periods still presents a challenge for MD. As a result, many multiscale coarse-grained MD
methods have been proposed in the past. These methods can be generally divided into
two classes. The first is based on coevolving both the atomistic and CG states
simultaneously \cite{Khuloud2002,Cheluvaraja2010,Long2010,Joshi2011,AbiMansour2014}; in these methods,
the atomic resolution is never lost. The second
class coarse-grains the atomistic structure and evolves
the CG variables based on dynamical equations that depend on parameters to be determined \cite{Bahar1997,Haliloglu1997,Chirikjian2005,Marrink2004,Murtola2007,Voth2008}.
These parameters are usually calibrated from experimental or MD simulation data.

In this work, we present \emph{ProtoMD}, a multiscale toolkit for prototyping both classes of multiscale MD algorithms.
\emph{ProtoMD} can be used as a program to run multiscale simulations or as a library to debug a multiscale algorithm or
perform coarse-grained analysis. \emph{ProtoMD} uses GROMACS \cite{GROMACS1995} to perform energy minimization and run MD simulations. This is needed for
completing the atomistic phase in coevolution methods and for calibrating system-specific parameters for CG-MD methods.

An overview of the theoretical methods on which \emph{ProtoMD} is based is presented in sec. (\ref{sec:Theoretical-Framework}), a discussion on the implementation of these methods and the design of \emph{ProtoMD} are provided in sec. (\ref{sec:Design-and-Implementation}), and a general user guide is covered in sec. (\ref{sec:User-Guide}).

%% file: Theoretical_Framework.tex
\section{Theoretical framework\label{sec:Theoretical-Framework}}

While \emph{ProtoMD} is a general-purpose toolkit that can be
used to prototype CG-MD methods (if calibration is based on data
collected from MD simulations), here we emphasize the implementation of multiscale
coevolution algorithms using the space-warping method \cite{Khuloud2002,Joshi2012}
as a coarse-graining technique.

Conventional MD is designed to solve Newton's equations for a set
of $N$ atoms interacting via conventional or custom force fields.
Coevolution methods take advantage of exntensive investment made
in the computational and software engineering of conventional MD codes.
This is achieved by integrating
MD into one complete multiscale algorithm that evolves an atomistic
system via MD, followed by coarse-grained dynamics. Coevolution methods have many of the
same goals as conventional MD, primarily to simulate the evolving
state of the $N$ atoms with atomic precision. However, these methods are able to
achieve this with greater efficiency in two distinct ways. The first
follows the evolution of an $N$ atom system along a time course that
starts with user specified initial data. The second is the analogue of ensemble
MD where hundreds of MD simulations are followed to provide
an assessment of the statistical significance of the simulation. These two types of
simulations (single and ensemble) are achieved via two distinct
algorithms implemented as two branches of the code.

Many-atom systems display cooperative behaviors ranging from mass
and charge density oscillations to structural transitions between
crystal phases in a polymorphic system. One challenge in many-particle
physics is to develop a quantitative understanding about how such
behaviors emerge from the Newtonian dynamics of many particles influencing
each other via an interatomic force field such as CHARMM \cite{charmm27} or AMBER \cite{amber1995}.
The standard statistical mechanical framework used to address such
a challenge is the Liouville equation (LE) for the dynamics of the $N$-particle
density, $\rho(\Gamma,t)$, for the $6N$ particle positions and momenta,
$\Gamma$. The pathway to achieving this quantitative understanding
considered here is to unfold the dependencies of $\rho$ on the microstate,
i.e., on both $\Gamma$ and a set of coarse-grained variables, $\Phi$.
Since $\Phi$ is constructed from $\Gamma$, then $\Phi$ obeys the equation of
motion $\dot{\Phi}=\Pi$, where $\Pi$ can also be expressed in terms of $\Gamma$ through Newton's
equations. With this, and the usual arguments of probability conservation
used to derive the LE, one finds that $\rho(\Gamma,\Phi,t)$ satisfies
the LE. Thus, an extended description of
the system state is introduced such that it includes the microstate (described
by $\Gamma$), and the mesostate (described by $\Phi$).
At first, this seems like a backward step in the quest for dimensionality
reduction. However, this expanded representation facilitates theoretical
developments and an associated conceptual picture, which ultimately
imply efficient and accurate computational algorithms. This extended
description introduces approximations that enter only via well-founded multiscale
perturbation \cite{Ortoleva2005,Ortoleva2009,Cheluvaraja2010,Joshi2011}
and Trotter factorization \cite{AbiMansour2014,Sereda2014}
methods. Besides the computational reduction, there are several advantages
to our approach. Key features of the $6N$ dimensional representation
are provided, and codevelopment of a multiscale perturbation technique
with the Trotter scheme yields insights into the physical meaning
of the latter. Both methods enable the use of well-established interatomic
force fields directly, avoiding the need to conjecture the form of
CG governing equations.

\subsubsection{Multiscale factorization \label{sub:Multiscale-Factorization}}

This technique is employed for the generation of a single $N-$atom
trajectory while an $N-$atom system simultaneously evolves on multiple
scales. This introduces a computational challenge for a time stepping
algorithm, i.e. capturing the slow modes as well as rapidly fluctuating
ones restricts the timestep. The latter determine the time stepping
of conventional MD integrators. However, the characteristic time of
the collective modes can be orders of magnitude larger than that of
atomic fluctuations. Since the collective behaviors are a result of
many individual atomic motions, the collective and fast-atom modes
are coupled, leading to the slowness of conventional MD. This is overcome
in multiscale factorization. Rather than attempting to untangle
the collective and fast particle modes directly, a Trotter
factorization strategy is used \cite{Trotter1959}.
In this approach, the
untangling of the collective and fast-atom modes is achieved via an
alternating stepping evolution procedure \cite{AbiMansour2014,Sereda2014}.
In each step, the collective and fast-particle modes are updated:
the former via the collective integration of the momenta constructed
from the MD phase, and the latter by conventional MD. This is justified via the stationarity hypothesis,
which states that the time evolution of the momenta conjugate to the collective variables can be
represented by a stationary process which expresses a representative ensemble of values
over a period of time short relative to the characteristic CG timescale \cite{AbiMansour2014}.
Then if the characteristic timescale of the collective
behavior is $\Delta$, and the MD duration needed to generate the representative ensemble
of conjugate momenta states is $\delta$, the theoretical efficiency over MD is expected to be $\Delta/\delta$.

\subsubsection{Multiscale perturbation \label{sub:Multiscale-Perturbation}}
This method generates an ensemble of $N$-atom trajectories and is therefore useful as an accelerated trajectory sampling technique, i.e. ensemble MD. Starting from the Liouville equation, Langevin type equations for a variety of CG variables are derived \cite{Ortoleva2005,Pankavich2008,Ortoleva2009}. The multiscale approach also provides prescriptions for constructing all factors (such as diffusion coefficients and thermal-average forces) in these CG governing equations, criteria for completeness of the set of CG variables, and algorithms for efficient simulation of $N$-atom systems based on coevolution of CG variables and an ensemble of all-atom configurations evolving with them \cite{Cheluvaraja2010,Joshi2011,Singharoy2011,Ortoleva2012}. The theoretical speedup over ensemble MD in this case is also $\Delta/\delta$, with $\delta$ being the length of the short MD runs constituting the microscopic ensemble needed to advance the mesoscopic state in time over a period of $\Delta$.

\subsection{Coarse graining and subsystems\label{sub:Coarse-Graining-and-ss}}

\begin{figure}
\centering
\begin{subfigure}{.5\textwidth}
  \centering
  \includegraphics[width=0.8\textwidth,natwidth=610,natheight=642]{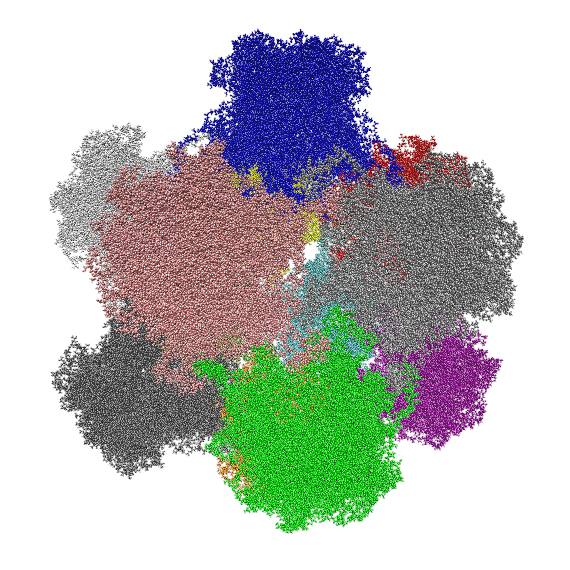}
\end{subfigure}%
\begin{subfigure}{.5\textwidth}
  \centering
  \includegraphics[width=0.8\textwidth,natwidth=610,natheight=642]{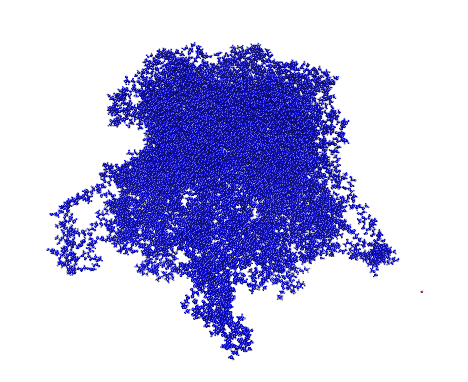}
\end{subfigure}
\caption{The capsid-like structure of L1 Human Papilimano virus (HPV) proteins (left)
is divided into twelve pentameric subsystems (right), each of which has its own
set of CG variables when subsystem decomposition is used in \emph{ProtoMD}.}
\label{fig:HPV}
\end{figure}

Interatomic forces in many-atom systems induce collective (i.e.,
coherent) behaviors which characterize the long space-time dynamics.
Identifying these collective variables is a first step in a multiscale
simulation. Mesoscopic systems also manifest single atomic modes (such
as collisions or vibrations) that are needed for a complete understanding
of the system behavior. While a variety of
CG variables have been introduced in the context of many-atom
simulations \cite{Clementi2011,Muller2002}, here the focus is on slowly varying
ones \cite{Khuloud2002,Pankavich2008,Joshi2012}
since they provide an appropriate starting point for multiscale analysis.
The general notion is to consider a set of CG variables $\phi$,
related to the $N$ atom positions $r$, via a dimensionality reduction
map. Complex systems such as viruses and other macromolecular assemblies
can often be separated into a collection of subsystems. Therefore,
it is convenient to apply this dimensionality reduction locally to each subsystem of the
entire structure. In the subsystem decomposition approach, subsystem $l$ has its own set of CG variables,
denoted $\phi_{l}$. For example, a virus capsid can often be decomposed
into energy-defined capsomers, notably pentamers or hexamers of a
protein (Fig. (\ref{fig:HPV})).

Specifically, let the center of mass of subsystem $l$ be denoted $R_{l}$ and
the subsystem in which atom $i$ resides be $l(i)$. Let $r_{i}$
be the position of atom $i$ and $s_{i}$ its position relative to
$R_{l(i)}$, i.e.,
\begin{equation}
r_{i}=R_{l(i)}+s_{i}.
\end{equation}
Using this notation, one can introduce CG variables describing the
position, orientation, and deformation of each subsystem. For example,
using the space warping method \cite{Khuloud2002,Joshi2012}, basis
functions $U_{kli}$ are introduced for atom $i$ in subsystem $l$,
where $k$ is the order of the polynomial spatial function used in
defining the CG variables. For example, if Legendre polynomials are
used in constructing $U_{kli}$, then each subsystem is embedded in a cube that contains the
subsystem and within which Legendre polynomials are used to define
the CG variables in terms of a reference atomic-configuration \cite{Singharoy2011}.
Let $\Phi_{kl}$ be the $k$$^{th}$ CG variable for subsystem $l$,
then the CG to all-atom map is taken to be a Fourier-like expansion
\begin{equation}
s_{i}=\sum_{k}\Phi_{kl}U_{kli}+\sigma_{i}.\label{eq:sum}
\end{equation}
The $k-$sum on the RHS describes the more coherent (i.e. collective)
dynamics of the system while the residual $\sigma_{i}$ accounts for
the small scale, stochastic motion of atom $i$ over-and-above the
coherent motion. The basis functions $U_{kli}$ are taken to be orthogonal
polynomials. The specific relation between $\Phi_{kl}$
and the atomic positions is determined by maximizing the information
in $\Phi_{kl}$. This is accomplished by minimizing the mass-weighted
sum of square residuals \cite{Pankavich2008}, restricted to atoms
in the given subsystem $l$.

As system complexity (e.g., the number of subsystems and their internal
structure) increases, one may increase the number of CG variables,
e.g., the range of the sum in Eq. (\ref{eq:sum}). As the number of CG variables
increases, smaller scale features are captured, but the characteristic
timescale decreases. This restricts the timestep used in advancing
the CG state. Too coarse a description might miss a given pathway
of, e.g., self-assembly or disassembly. Thus, there is a trade-off
between completeness and numerical efficiency. However, this strategy
has been shown to be successful in arriving at an accurate and efficient
algorithm for simulating bionanosystems \cite{Singharoy2011,Joshi2012,Jing2014}
since the CG variables guide the overall dynamics, while much of the atomic-scale
structure is captured by the fine-scale aspect of the coevolution.

The implementation of both multiscale algorithms and the subsystem decomposition outlined here
is discussed in the next section that covers the technical aspects of the design of \emph{ProtoMD}.

%% file: Design_and_Implementation.tex
\section{Design and implementation\label{sec:Design-and-Implementation}}

\subsection{Strategy}

One of the key goals in the design of \emph{ProtoMD} is to allow users to readily
experiment and develop their own CG variable definitions and dynamical
equations. The overall structure of \emph{ProtoMD} is based on an object-oriented
approach, with strict enforcement that all functions are guaranteed
to be side effect free. Side effect producing functions can modify
global variables or input parameters without the user being aware
of it. Such side effects are eliminated with the approach adopted
here, which proved to be practical when dealing with large sets of numerical data
(atomic coordinates, velocities, etc., for
supramillion atom systems). All state management is isolated in the top level
System class. The \emph{System} object contains a number of other objects.
However, once these objects are initialized, they are immutable. For example,
the integrators do not have any internal mutable state, i.e., they simply
take an existing atomic state (from the System.current\_timestep context)
and calculate the next state.

\subsection{Functions and classes\label{sub:Functions}}
A system is a set of atoms that can be delineated
in a number of ways. We specify subsystem delineation with an MDAnalysis
selection string. MDAnalysis provides a full atomic selection language
comparable to VMD \cite{VMD} or CHARMM \cite{CHARMM1983}.
Subsystem CG variables are handled with the
\emph{proto\_md.subsystems.SubSystem} class. A \emph{SubSystem} references
a specific set of atoms, calculates
its associated CG variables, or conversely it can take new CG variables
and construct the micro state, or an ensemble thereof, consistent
with the new CG state. SubSystems are notified by the System of various
events such as when the energy of the atomic state is minimized, the system is equilibrated,
or connectivity has changed. The atoms contained in a \emph{SubSystem} are stored
in an MDAnalysis AtomGroup object, i.e. each \emph{SubSystem} contains an \emph{AtomGroup}.

\emph{SubSystems} are created by a factory function. The factory creates objects
without specifying the exact class of object that will be created.
The essence of this pattern is to define an interface for creating an object,
and let the classes that implement the interface decide which object to instantiate.
Therefore, the factory method lets a class defer instantiation to subclasses.

A \emph{SubSystem} factory function must have a signature of \emph{def SubsystemFactory(system, selects, {*}args)},
where \emph{system}
is a reference to the System object, \emph{selects} is a list of MDAnalysis
\emph{Atomselect} statements, and \emph{args} is a list of arbitrary user specified
data. The full textual name of the subsystem factory function is stored
in the database, e.g. \emph{proto\_md.subsystems.RigidSubsystemFactory}.
The System object's initializer dynamically instantiates an instance
of the named factory function and calls it to create a list of \emph{SubSystems}.
This approach to object creation maximizes the flexibility and the
ease with which one can extend \emph{ProtoMD}. All one needs to do is provide
the textual name of a factory function. This named function does
not need to be part of \emph{ProtoMD}. The named factory function simply needs
to be reference-able in the python path. So this approach allows one
to extend \emph{ProtoMD} without having to change any \emph{ProtoMD} source code. This
could prove to be very useful in a situation where one is developing some
new type of CG variable and one does not have administrator privileges,
and \emph{ProtoMD} is installed in a system path. Users can develop a new \emph{SubSystem}
object, store it in their home directory, and simply specify the name
of their subsystem factory in the \emph{ProtoMD} database; \emph{ProtoMD} will automatically
use their new \emph{SubSystem} object.

A \emph{SubSystem} derived object must implement the \emph{SubSystem} interface.
Because the implementation is in python, it is technically not a requirement for subsystems
to be derived from the \emph{SubSystem} class; they simply must define
and implement all the methods defined by the \emph{SubSystem} class. The
\emph{proto\_md.subsystems.SubSystem} class only defines the interface that subsystem
classes must implement.

\begin{itemize}
\item \emph{def universe\_changed(self, universe):} changes the universe.
This occurs when the number of atoms has changed (on startup, in the
event of solvation or chemical reactions). So, \_\_init\_\_ will
be called when a new \emph{SubSystem} derived object is created,
then after the universe is created, it will call universe\_changed.
\item \emph{def frame(self):} notifies the subsystem that a new frame is
ready. It should return a tuple (position, velocity, force) of CG variables
for the current frame. If the CG positions are stored in a multi-dimensional
tensor (e.g., $3\times3\times3$), then they need to be unfolded and
returned as a $1 \times 27$ row vector, and similarly for the CG velocity
and force variables.
\item \emph{def translate(self, values):} translates all atomic positions
given a coarse grained variable. For example, if the CG variable is the
center of mass of a macromolecule, then translate would be given a $1 \times 3$ vector which represents
how much to shift the center of mass. Next, the subsystem would simply
add this value to all atomic positions. The \emph{values} argument is the
difference in CG variable space between the current configuration
and a deformed configuration. This is given as an unfolded $1\times n$
column vector, and if the CG variables are tensorial, then \emph{values}
needs to be folded.
\item \emph{def minimized(self):} is called after the atomic system is minimized;
no return value is required. This method is provided if the CG variables must be modified
in the event of minimization.
\item \emph{def equilibrated(self):} is called after the atomic system is
relaxed (thermalized in isothermal simulations); no return value is required.
\item \emph{def md(self):} is called after the MD trajectories are processed; no return value is required.
\end{itemize}
\emph{ProtoMD} currently provides two types of SubSystems. The first uses the
center of mass as the CG variable and is denoted \emph{proto\_md.subsystems.RigidSubsystem}. This subsystem serves as a demonstration,
because it is probably not useful beyond spherically symmetric systems. A more sophisticated and useful sybsystem is \emph{proto\_md.subsystems.SpaceWarpingSubSystem},
which constructs a set of CG variables based on an orthogonal polynomial basis set \cite{Khuloud2002,Joshi2012}. In principle,
any set of orthogonal polynomials may be used, but currently only Legendre polynomials are implemented.

\subsection{Coevolution integration\label{sub:Time_Integration}}

The \emph{Integrator} class performs the time evolution of the system with the
application of the classical propagator $\Gamma(t_{n+1})=e^{i\mathcal{L}t}\Gamma(t_{n})$.
The \emph{Integrator} currently has two concrete implementations actualizing
the physical ideas presented in sec. (\ref{sec:Theoretical-Framework}): the LangevinIntegrator
implements the concepts presented in sec. (\ref{sub:Multiscale-Factorization})
while the FactorizationIntegrator implements those presented in
sec. (\ref{sub:Multiscale-Perturbation}).

\subsection{Key objects and packages}

\begin{figure}
\centering{}\includegraphics{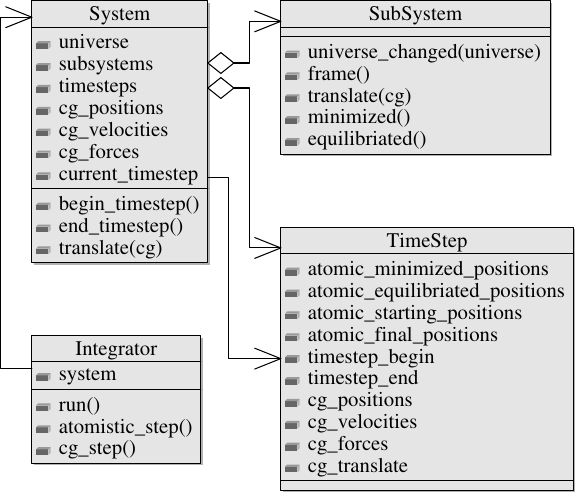}\caption{A UML diagram of the key objects and their key attributes in \emph{ProtoMD}.\label{fig:UML-diagram}}
\end{figure}

\emph{ProtoMD} consists of a number objects, the key constituents are discussed
here, and are displayed in Fig. (\ref{fig:UML-diagram}).

The principle object is conveniently called the \emph{System}.
This object manages the state of the simulation and is roughly analogous
to OpenGL context (Fig. (\ref{fig:openGL})). All state variables are contained entirely in
the System. Additionally, all other objects are contained within the
System.

\subparagraph{system.System}

system.System is the top level class, it is a container class for
all the other classes. It is also the single place where \emph{ProtoMD} handles
data persistence. The System class can be used either for loading
and simulation, or loading and analysis. The System constructor takes
the path to a database file, and a flag indicating the mode to open
the file, defaults to ``r'' - read only. For running a simulation,
it should be ``a''. This flag is relevant in that a simulation
can be running in one process (in read / write mode), and another
process can safely open the database in read only mode and not interfere
with a running simulation. This is useful for checking the state
of a running simulation.

All simulation data is stored in a single database file, and the System
class provides a programatic interface. All completed timesteps are
accessible via the timesteps property.

\subsubsection{State management and data persistence}

The present state of a running simulation is contained wholly within
the System.current\_timestep and System.universe objects.

Molecular simulation programs have typically stored their data in
a variety of proprietary, non-standardized, and incompatible file formats.
For example, GROMACS \cite{GROMACS2005} uses 31 different file formats, some of which
are deprecated or used only as input. Even though most MD file formats are documented, accessing them entails
developing a parser for each format. Rather than invent yet another
format, \emph{ProtoMD} used the standardized and widely supported hierarchical
data format version 5 (HDF). HDF is a self-describing file format
designed for the general exchange of numerical data.

All information relevant to a \emph{ProtoMD} simulation is stored in a single
HDF5 database. This approach allows preparation of a database on a
completely different computer than the one on which the simulation
is intended to be run. All molecular dynamics topologies, force field
parameters, simulation parameters, coordinates, etc., are stored in
the HDF5 database. One could even start a simulation on one compute
cluster, run for a while, and restart the simulation on a different cluster
by simply copying a single file.

Considering the environment in which MD simulations are run, a compute
cluster is sometimes unreliable. Compute nodes and
data storage frequently crash or are rebooted. Programs are often
terminated by the job scheduler. Flaws or faults in external programs
become evident, and countless other environmental conditions can cause
termination of executing programs. Molecular dynamics simulations
produce a considerable amount of data. \emph{ProtoMD} also produces and uses
a good deal of disparate data, ranging from simulation parameters,
archived molecular dynamics input files, and atomic trajectory
time series. Therefore, an efficient, fault-tolerant data storage
system which readily allows restarts is a key, if not the core, requirement.

For on disk file storage, the HDF5 file format was chosen. HDF stands
for ``hierarchical data format.'' HDF5 is an open standard file
format for the high performance storage and retrieval of an unlimited
variety of datatypes. HDF5 provides a library that allows access in
a variety of languages. In this sense, it can be thought of as similar
to XML. However, XML is plain text (though it can be compressed),
it is nonetheless inefficient for storage of large amounts
of numerical data. HDF5 can can be accessed natively in a variety
of different mathematical and analysis packages such as Mathematica,
MATLAB, and R.

A data-persistence model was chosen that is similar to that of most graphics
libraries such as Cairo, OpenGl, Cocoa, or Windows. All of these libraries
function essentially the same way. One creates or obtains a graphics
context for an appropriate output device, be it the screen, a printer,
or an image buffer. The graphics context is a data structure which
maintains information about the current drawing state. Then there
are drawing functions, such as drawRect, drawPoly, and so forth, which
add graphics primitives to the context. In the case of OpenGl, one
adds a sequence of 3D information to the context.
The key point here is that all of these drawing operations operate
on the data structure, and not the output device directly. Once all
of the drawing operations are complete, then the graphics context
is passed to the output device in a single atomic operation. In contrast, if one
were to output each drawing operation immediately \textbf{to} the
output device, it would be very cost prohibitive as that output device
must be locked, data written to it, and then the device must be unlocked
again. It would be a similar case if every function that produced
some relevant output information wrote that information directly to
disk and flushed the IO buffers. Therefore, all operations that affect
the state of a TimeStep write to the TimeStep data structure, which
is a memory resident object. The TimeStep object is created by the
\texttt{System.begin\_timestep()}, accessed with the \texttt{System.current\_timestep},
and atomically persisted to disk and deleted from memory via\texttt{
System.end\_timestep()}.

\begin{figure}
\begin{lstlisting}[language=C, showstringspaces=false]
void display() {
	glBegin(GL_LINES);
	glVertex3f(1.0, 1.0, 0.0);
	glVertex3f(-1.0, -1.0, 0.0);
	glEnd();
	glFlush();
\end{lstlisting}
\caption{A sample OpenGL code that displays a line on the screen. Objects in \emph{ProtoMD} manage the state of the simulation in a similar way
to that employed in OpenGL.}
\label{fig:openGL}
\end{figure}

The TimeStep object behaves as a data structure, however,
it is actually implemented as an HDF5 data set. The reason for this
is that the data stored in a TimeStep object can potentially be huge.
Therefore, the underlying HDF5 implementation determines
how to best store this data, e.g., all data directly in memory, or some
portions stored on disk.

\subsection{Simulation}

The command-line interface of \emph{ProtoMD} can be accessed via \emph{python -m proto\_md}.
This front end program is intended for initial database generation
and running the simulation. There is an extensive built in help system
accessible via: \emph{python -m proto\_md --help}. Available options will
likely change in the future, so it is best to consult the help system
for the latest complete up-to-date information.

This subsection will describe how a simulation is performed from both
implementation and user perspectives.

\subsubsection{Initial database creation}

The first step in setting up a simulation is the creation of a database.
We have provided a command line program that front ends the config.create\_sim(...)
function. This function (program) reads a set of parameters such as
simulation parameters, subsystem definition and atomic structure files.
It can optionally generate a topology (using pdb2gmx), perform a series
of validations, and finally package everything up into a \emph{ProtoMD} database
file, from which a \emph{ProtoMD} simulation can be run.

\subsubsection{Stand-alone programs and libraries}

\emph{ProtoMD} can be used either as a stand alone program, or as a library.
There are many instances where it is more convenient to use \emph{ProtoMD} as
a program, e.g. submission of simulations to a queuing system, or
quick and simple analysis tasks. The stand alone program interface
to \emph{ProtoMD} is made possible by python's module mechanism, i.e. when a
package is installed with a \texttt{\_\_main\_\_.py} file.

\subsubsection{Interfacing with molecular dynamics programs}

\emph{ProtoMD} requires external MD programs to perform a number of
calculations, and it interfaces with external MD programs using a very transparent
approach; \emph{ProtoMD} handles all interaction with external programs through
\emph{context managers}. All of the MD related functionality is contained
in the \texttt{proto\_md.md} package.

Programmatically, interfacing with external MD programs is frequently
fraught with error primarily due to the large number of input files
that these MD programs require to be present on the file system. Most
MD packages do not provide an application program interface (API), and are
intended to be user-invoked programs. This presents difficulty to
calling programs, as they must determine the current state of the
file system, copy and convert large portions of internal state into
MD input files, call the MD program, parse the output, and finally
clean up after the MD program is run. Significant sources of error
arise when the MD program reads state from locations that the calling
program is unaware of, for example, external configuration files,
and force field parameter files. Furthermore, a more common source
of error is the MD program reading previous input files that perhaps
were not deleted properly. In effect, most of these errors arise from
the fact that input state required by MD can be scattered throughout
the file system. This can be analogous to a programming language with
no variable scoping, i.e. all variables are global: in such situations,
one loses the concept of functions, and all subroutine calls consist
purely of side-effects. Yet another error occurs when concurrent copies
of the calling program are run: if these programs have to use the
file system to pass state to an MD program, they will overwrite each
others' file system state.

To circumvent or  sequester these errors,
\emph{ProtoMD} generates temporary, randomly named directories into which the
requisite MD input state variables are serialized, the necessary files
are copied, and various GROMACS programs are run via \emph{GROMACS wrapper} \cite{Beckstein}.
The use of temporary randomly named directories to store
MD state minimizes the possibility of using other files or input that the
calling program is unaware of. These temporary
directories only exist for the lifetime of the MD program call. In
effect, we have created \emph{stack frame}, the set of information
required for, and existing only for the lifetime of procedure call.

The lifetime management of these temporary file system based variables
is implemented using a Resource Acquisition Is Initialization
(RAII) idiom \cite{Stroustrup1994}. This is a programming idiom, widely attributed to Bjarne
Stroustrup, which states that a class with a constructor that
acquires the resource (either by creating it, or obtaining it via a parameter)
has a destructor that always releases the allocated resource. During
the object lifetime, the resource may be accessed via instance variables,
and is guaranteed to be freed when the object is either deleted or
goes out of scope. RAII is a widely used pattern in C++: it is a safe
way to deal with resources and also makes the code much cleaner as
it eliminates the need to mix error handling code with functionality.
This idiom was traditionally not used in non-deterministic, garbage
collected languages like python. In C++, the object is destroyed when
it either goes out of scope (if stack allocated), or deleted (if heap
allocated). In python, however, there is no explicit control over
when the object is destroyed. Therefore, if it contained a resource,
that resource may potentially exist for the lifetime of the program.

This limitation was overcome with PEP343, where \emph{context managers},
provide \texttt{\_\_enter\_\_()} and \texttt{\_\_exit\_\_()}
methods that are explicitly invoked on entry to and exit from the
body of the \texttt{with} statement. For instance, the open statement
returns a context manager containing an open file handle, it is kept
open as long as the execution is in the context of the \texttt{with}
statement where it is used, and closed when execution leaves the block,
irrespective of whether the exit was because of an exception or during regular
control flow. The \texttt{with} statement can thus be used in ways
similar to the RAII pattern in C++: some resource is acquired by the
\texttt{with} statement and released when the \texttt{with} block
is left.

\emph{ProtoMD} wraps all external MD calculations into a set of functions
which return a \texttt{proto\_md.md.MDManager} object. This is a context
manager which holds all of the file system objects required for, and
returned by, MD programs in the file system directory. Calling functions
can access any of the MD results through a dictionary interface. For
example, the function \texttt{md.solvate(...)} accepts an
\texttt{MDAnalysis.Universe} object, along with other parameters,
creates a random temporary directory, write the contents of the universe
object, performs a solvation using external GROMACS programs,
and finally returns a context manager containing this directory with
the solvated structure and topology files. Calling functions can access
these files with the ``struct'' and ``top'' keys, i.e.
\begin{lstlisting}[language=Python, showstringspaces=false]
with md.solvate(...) as sol:
	do_something(sol["struct"], sol["top"])
\end{lstlisting}
As soon as the \texttt{with} block is exited, the context manager
deletes the temporary directory, and all its contents. Therefore,
we have eliminated the chance that subsequent calls to external MD
programs could pick up input files that were left in this directory.
Furthermore, as each call to an MD function creates a unique and random
directory, multiple copies of \emph{ProtoMD} may be concurrently executed.

\subsubsection{Solvation}

The CG variables used in \emph{ProtoMD} are functions of
atomic coordinates of the macromolecule of interest (although in principle, one could define
the solvent as a subsystem). In MD simulations, the solvent atoms
typically account for at least 50\% to 75\% by particle count of the
total system. After the end of each CG timestep, \emph{ProtoMD} extracts the macromolecule,
which is then resolvated using Gromacs, and counter-ions (by default NaCl) are added to the system. The reason
resolvation is done is to avoid bad atom contacts between the macromolecule and the solvent. In the case of simulations
done in vacuum or using implicit solvent models, solvation is omitted.

%% file: User_Guide.tex
\section{User guide\label{sec:User-Guide}}

The purpose of this section is to describe how users can launch and
analyze multiscale simulations via the \emph{ProtoMD} platform.
\emph{ProtoMD} can be used as an executable module, i.e. a program with
command-line arguments. In this mode, a GROMACS compatible structure
input file can be used to initiate a multiscale simulation. The user
must also supply GROMACS-specific parameters for running MD simulations
(used for the microscopic phase), and \emph{ProtoMD}-specific parameters needed
for the mesoscopic phase wherein the CG variables are advanced in time and new
microstates are generated. The output is an hdf file that contains
information including the trajectory at user-specified CG time steps. In
the second mode, \emph{ProtoMD} can be used as a library containing modules that
can be assembled into programs; this mode facilitates debugging runs
and analyzing results.

\subsection{Genaral concepts}

\emph{ProtoMD} is built on the concept that everything  related to a simulation
(including configuration parameters and output)
is stored in a single database file. Most traditional MD programs rely
on a large number of configuration and parameter files. Even though having a number of
files in different locations makes it relatively easy for users to
modify these files, it is more convenient to reproduce a single input file when rerunning a simulation
at either a later date or on another computer.

Since \emph{ProtoMD} is a wrapper around GROMACS, it requires GROMACS-specific
files. When a simulation is first prepared, all these external GROMACS
configuration files are collected and stored in the \emph{ProtoMD} database file.
These files are automatically extracted and passed to GROMACS when
the simulation is run. This ensures that results are always reproducible
because there are no external dependencies. Furthermore, this approach
facilitates moving or restarting simulations on different computers.
Subsection \ref{sub:Preparing-a-System} shows how to use the
\emph{ProtoMD} command line interface to prepare an initial database.

One can iterate through the output trajectory frames in a very similar
manner to that used in MDAnalysis \cite{Beckstein2011}. Details of how one can programmatically
access the \emph{ProtoMD} database including trajectory analysis is covered in
section \ref{sub:Acessing-the-databse}.

Nearly all \emph{ProtoMD} objects and methods have descriptive docstrings.
These strings form the contents of the interactive python help system,
and it can be accessed by typing the name of the class or method followed
by a ``?'' in the interactive python interpreter.

\subsection{Selections}

A key concept in \emph{ProtoMD} is atomic selections. Atomic selections are a
group or subset of atoms selected from the whole system based on some
criteria. They are used throughout \emph{ProtoMD}, particularly in defining subsystems (sec. \ref{sub:Coarse-Graining-and-ss}).
MDAnalysis \cite{Beckstein2011} provides a selection language
with a syntax similar to that of CHARMM \cite{CHARMM1983} or VMD \cite{VMD}. One can
think of this atomic selection language as ``SQL for atoms''. The
atomic selection language provides a way to select atoms based on
a variety of criteria such as boolean combinations of simple keywords,
geometric or connectivity measures, index, name, residue ID, or segment
names. Examples on selection statements using MDAnalysis are as follows:

\begin{lstlisting}[language=bash,tabsize=4, showstringspaces=false]
# select all DMPC lipid atoms
universe.selectAtoms("segid DMPC")
# select all atoms within 5.0 Angstroms from (0,0,0)
universe.selectAtoms("point 0.0 0.0 0.0 5.0")
# select all atoms belonging to residues 1 to 10
universe.selectAtoms("resid 1:10")
\end{lstlisting}

In \emph{ProtoMD}, the user creates a list of selections to define the subsystems.
These selections must not overlap, however, because each subsystem
must constitute one or more segment(s). For example,
it is natural to define the subsystems at the pentamer or hexamer
level for a virus-like particle (Fig. (\ref{fig:HPV})).

\subsection{Preparing a system\label{sub:Preparing-a-System}}

\emph{ProtoMD} can be run from the command line as follows: python -m proto\_md command
{[}args{]}, where command can be one of the following:
\begin{itemize}
\item config: creates a database from which a multiscale simulation can
be run
\item top: auto-generates a topology file
\item solvate: attempts auto-solvation
\item run: runs or resumes a simulation
\item runsol: like run, but performs only the auto-solvation step
\item mn: performs energy minimization
\item mneq: performs energy minimization followed by equilibriation
\item mneqmd: performs energy minimization, equilibriation, and molecular
dynamics
\item eq: performs equilibriation
\item atomistic\_step: performs a full atomistic step (microphase)
\item step: performs a single complete coarse-grained step (mesophase)
\item md: performs only an MD step with the starting structure
\item cg\_step: performs only the coarse grained portion of the time step
\end{itemize}

\subsubsection{Configuring the system}

The first step in running a multiscale simulation is to create the
initial database. This can be accomplished by using the following
commands: python -m proto\_md config {[}args{]} where args can be one of
the following:
\begin{itemize}
\item -o fid FID name of the simulation file to be created.
\item -struct STRUCT starting structure name.
\item -box BOX BOX BOX x,y,z values of the system box in Angstroms. If box
is not given, the system size is read from the CRYST line in the structure
pdb.
\item -top TOP name of the topology file. If this is not given, \emph{ProtoMD} attempts
to generate a topology file using pdb2gmx.
\item -posres POSRES name of a position restraints file, optional.
\item -I INCLUDE\_DIRS name of the directories to search for topology file includes.
There can be many additional include directories, just like gcc, but
UNLIKE GCC, there must be a space between the -I and the dir, for
example -I /home/foo -I /home/foo/bar.
\item -temperature TEMPERATURE is the temperature (in Kelvin) at which to run the simulation. The default temperature is $300$ K.
\item -subsystem\_factory SUBSYSTEM\_FACTORY is a fully qualified function name
which can create a list of subsystems. The function defaults to \emph{proto\_md.subsystems.SpaceWarpingSubsystemFactory}
which uses the space warping method for constructing the CG variables
for each subsystem. Currently, only Legendre polynomials are supported.
\item -subsystem\_selects SUBSYSTEM\_SELECTS {[}SUBSYSTEM\_SELECTS ...{]}
is a list of MDAnalysis select statements, one for each subsystem. For
example, if the system is divided into 10 subsystems, subsystem\_selects
takes a list of length 10, with each entry in the list defining a
given subsystem.
\item -anion anion name to use for ionization (if system is not neutral). The default anion is Cl$^-$.
\item -cation cation name to use for ionization (if system is not neutral). The default cation is Na$^+$.
\item -concentration concentration (in mol/L) of ions to use for ionization. The default is $0.15$ M.
\item -subsystem\_args SUBSYSTEM\_ARGS {[}SUBSYSTEM\_ARGS ...{]} is a list
of additional arguments passed to the subsystem factory, the second
item of the list may be the string \emph{resid unique}, which creates a
separate subsystem for each residue. The most commonly used subsystem
is \emph{SpaceWarpingSubsystemFactory}. The args for this subsystem
are {[}kmax, OPTIONAL(``resid unique''){]}. kmax provides the upper
bound on the sum of the three Legendre indices, and the last argument
is the optional string ``resid unique'' to make a unique subsystem
for each residue.
\item -integrator INTEGRATOR is a fully qualified name of the integrator function,
defaults to \emph{proto\_md.integrators.LangevinIntegrator}, and the other integrator
provided is \emph{proto\_md.integrators.FactorizationIntegrator}.
\item -integrator\_args INTEGRATOR\_ARGS are additional arguments passed to
the integrator function.
\item -cg\_steps CG\_STEPS is the number of coarse grained time steps taken
over the duration of a multiscale simulation.
\item -dt DT is the size of the coarse grained time step in picoseconds.
\item -mn\_steps MN\_STEPS is the number of MD steps used in performing energy minimization.
\item -md\_steps MD\_STEPS is the number of MD steps used for performing the MD
run of the microphase
\item -multi MULTI is the number of parallel MD runs used to construct microstate
ensembles for the Langevin integrator.
\item -mn\_args MN\_ARGS is a dictionary of GROMACS-specific (mdp) parameters used in minimization.
\item -md\_args MD\_ARGS is a dictionary of GROMACS-specific (mdp) parameters used in MD.
\item -ndx NDX is the name of GROMACS-specific index file.
\item -solvate is a boolean flag. If it is set and the system is to be auto-solvated, the initial structure
must not contain any solvent. This flag default to False. To enable solvation, add `-solvate' with no args.
\item -debug enables debug mode (all simulation directories are not deleted).
\item -mainselection MAINSELECTION is the name of make\_ndx group which is
used for the main selection. This should be a group that consists
of the non solvent molecules. ``Protein'' usually works when simulating
protein. However, another selection command is required when simulating
lipids. To find which selection to make, either run \emph{ProtoMD} config, and an error will
pop up informing the user of the available groups, or run make\_ndx to
get a list of possible selections.
\end{itemize}

\subsubsection{Examples}

Algorithms on which \emph{ProtoMD} is based has been demonstrated for a number of systems \cite{Cheluvaraja2010, AbiMansour2014, Jing2014, Sereda2014, AbiMansour2015}.
The following example illustrates the use of the \emph{ProtoMD} command line
interface to generate a simulation database. In the following example,
a database file named ``1LFG-open.hdf'' is created using the starting
structure of Lactoferrin, ``1LFG-open.gro''. \emph{ProtoMD} generates
a single-trajectory simulation (multi $1$) for a series of $30$ CG time steps,
each of length $1$ ps (dt 1). The order of Legendre polynomials to be used is less
than or equal to $1$ (subsystem\_args $1$), which implies a total of $4 \times 3$ CG variables are
used to coarse-grain the protein. Furthermore, this example generates
a single subsystem which is specified by the ``not segid SOL'' selection statement.
This statement specifies
that the subsystem should consist of all the atoms present in the
system which are not solvent atoms (i.e., protein atoms). Note that here no topology file
is specified; in this case, \emph{ProtoMD} uses the `pdb2gmx'' program to generate
a topology file, which is stored in the hdf database.

\begin{lstlisting}[float,language=bash,tabsize=4, showstringspaces=false]
#!/bin/sh
python -m proto_md config \
	   -o 1LFG-open.hdf \
	   -struct 1LFG-open.gro \
	   -temperature 300.0 \
	   -subsystem_selects "not segid SOL" \
	   -cg_steps 30 \
	   -dt 1 \
	   -mn_steps 5000 \
	   -md_steps 500 \
	   -multi 1 \
	   -integrator \
	    proto_md.integrators.FactorizationIntegrator \
	   -subsystem_factory \
	    proto_md.subsystems.SpaceWarpingSubsystemFactory \
	   -subsystem_args 1
\end{lstlisting}

\subsection{Accessing the database and analyzing simulations\label{sub:Acessing-the-databse}}

\emph{proto\_md.System} object is the main class in \emph{ProtoMD} responsible for loading
and accessing the underlying hdf5 database, and for orchestrating
the various MD programs used to advance the simulation.
The System object allows one to open the underlying hdf5 database and
access all pertinent data as native python objects. One can create
a System object in either read/write or read-only mode. One typically
creates a System object in read-only mode to analyze and access simulation
results, and only the actual running simulation creates the System
object in read/write mode. There can be multiple instances of the
System object being open in the same database. For example, a simulation
can be currently running (on another process or even another node
in a compute cluster), and a user can create a System object to
access the simulation. Provided the user creates the System object
in read-only mode, it will not interfere with the currently running
simulation.

The System object has a number of important properties and methods.
However, from an analysis perspective the most important is the System.timesteps
property. This is stored in the database and a series of
hdf5 groups, but is presented to the user as a list of Timestep objects.
The Timestep object provides access to all of the calculated variables
in a CG step as Numpy arrays, and the Timestep object can even create
an MDAnalysis Universe object from the CG timestep data.

The Timestep object provides the following properties:
\begin{itemize}
\item atomic\_minimized\_positions
\item atomic\_equilibriated\_positions
\item atomic\_starting\_positions
\item atomic\_final\_positions
\item timestep\_begin
\item timestep\_end
\item cg\_coords
\item cg\_velocities
\item cg\_forces
\item cg\_translate
\end{itemize}
The Timestep object is also capable of writing itself to the hdf5
database. This is accomplished with the Timestep.flush() method. Writing
to the database is normally only performed by a simulation, and will
fail if the System was opened in read-only mode.

For analysis, the Timestep object provides a useful ``create\_universe()''
method, which creates an MDAnalysis Universe from the timestep
object. Once the Universe is created, it can be loaded with
the atomic positions listed above. Syntactically, it would have been
simpler if each timestep possessed a universe property; however,
the implementation of such a system would have entailed either creating
a new universe each time the universe property is called (computationally
expensive), or create a caching system. It is relatively simple to
first obtain a universe object then set the atomic properties during
the course of iterating over the timesteps.

\subsubsection{Examples}

This section provides an example on analyzing a \emph{ProtoMD} trajectory using
the universe object and the timesteps.
\begin{itemize}
\item Radius Of Gyration: This example first obtains a universe object,
then sets the atomic positions with the values stored in each timestep.
Finally, it performs the calculation and appends the value to a list.

\singlespacing
\begin{lstlisting}[language=Python, showstringspaces=false]
# import the proto_md package
import proto_md

# create a system from the file `my_simulation.hdf'
sys = proto_md.System(`my_simulation.hdf')

# create a new universe object, this is now
# set to the initial coordinates
u = s.universe

# create a new empty list
l = list()

# iterate over the timesteps
for ts in sys.timesteps:
	# set the universe atomic positions
	# to the values stored in the database timestep.
	u.atoms.positions = ts.atomic_final_positions

	# calculate the radius of gyration
	rog = u.atoms.radiusOfGyrations()

	# append rog to the list
	l.append(rog)
\end{lstlisting}

\doublespacing
\item Radius Of Gyration (alternative version): perhaps a more
`pythonic' method of calculating the radius of gyration (or any other
metric) would be to create a higher order function once the universe
is obtained and populate it with the atomic positions and return it.
In this approach, list comprehension can be used to create a list
of values using a single list comprehension statement:

\singlespacing
\begin{lstlisting}[language=Python, showstringspaces=false]
# import the proto_md package
import proto_md

# create a system from the file `my_simulation.hdf'
sys = proto_md.System(`my_simulation.hdf')

# create a new universe object, this is now
# set to the initial coordinates
u = sys.universe

# function which takes an array of atomic positions,
# sets the universe atomic state and returns the
# universe object.
\end{lstlisting}
\begin{lstlisting}[language=Python, showstringspaces=false]
def uf(pos):
	# set the universe's atomic positions
	u.atoms.positions = pos
	return u

# calculate the value in a single list statement.
rog = [uf(ts.atomic_final_positions).atoms.radiusOfG-
yration() for ts in sys.timesteps]
\end{lstlisting}

\doublespacing

\item Distance Between Proteins: This example is taken directly from the
MDAnalysis website, but here MDAnalysis is used to analyze a coarse
grained time series. In this example, the end to end distance of a
protein and the radius of gyration are calculated. This is assuming
that the protein in question has the segment identifier ``S4AKE''.
This example uses MDAnalysis atom groups. Note, any analysis of \emph{ProtoMD}
timesteps performed using MDAnalisys virtually identical to conventional
MD analysis, the only difference being that one must to set the atomic
positions from the \emph{ProtoMD} timestep values.

\singlespacing
\begin{lstlisting}[language=Python, showstringspaces=false]
# import the proto_md package
import proto_md

# create a system from the file `mySim.hdf'
sys = proto_md.System(`mySim.hdf')

# create a new universe object, this is now
# set to the initial coordinates
u = s.universe

# the first nitrogen atom in the 4AKE segment.
nterm = u.s4AKE.N[0]

# last carbon atom in this segment
cterm = u.s4AKE.C[-1]

# a selection (a AtomGroup)
bb = u.selectAtoms(`protein and backbone')

# iterate through all frames
for ts in sys.timesteps:
	# set the universe atomic positions
	# this is the only different aspect from
	# conventional MD time series analysis.

	u.atoms.positions = ts.atomic_final_positions

	# end-to-end vector from atom positions
	r = cterm.pos - nterm.pos

	# end-to-end distance
	d = numpy.linalg.norm(r)

	# AtomGroup is a `live' object, it is updated
	# each time new coordinates are loaded.
	rgyr = bb.radiusOfGyration()

	print "frame = %d: d = %f Angstrom, Rgyr = " \
	      "%f Angstrom" % (ts.frame, d, rgyr)
\end{lstlisting}

\doublespacing

\end{itemize}

\subsection{Running simulations}

Once an hdf file is created, a multiscale simulation is initiated
by invoking: \emph{python -m proto\_md run file.hdf {[}-debug{]}}. If the user supplies the optional parameter -debug,
\emph{ProtoMD} will not delete
any of the directories created to run MD. This is useful for debugging
and tracking problems with unstable simulations.

As an example on running a multiscale simulation using the Trotter integrator, pertussis toxin (PDB ID: 1PRT), a protein-based AB5-type exotoxin \cite{1PRT}
was simulated under NVT conditions at $300$K. The CHARMM27 force field \cite{charmm27} was used in explicit solvent using the TIP3P model \cite{TIP3P}. NaCl counter-ions of concentration $0.15$ M were added for charge neutrality. The system consisted of $603,775$ atoms in a box of dimensions $16 nm \times 16 nm \times 24 nm$. The CG timestep was set to $0.5$ ps with the micro (MD) phase taken to be $100$ fs and the equilibriation time $50$ fs. To prepare the system, the following script was run:

\begin{lstlisting}[language=bash,tabsize=4, showstringspaces=false]
#!/bin/sh
python -m proto_md config \
	   -o 1PRT.hdf \
	   -struct 1PRT.gro \
	   -temperature 300.0 \
	   -subsystem_selects "protein" \
	   -cg_steps 7400 \
	   -dt 0.5 \
	   -mn_steps 0 \
	   -eq_steps 50 \
	   -md_steps 100 \
	   -top_args {'ff': 'charmm27'} \
	   -multi 1 \
	   -integrator \
	    proto_md.integrators.FactorizationIntegrator \
	   -solvate \
	   -subsystem_factory \
	    proto_md.subsystems.SpaceWarpingSubsystemFactory \
	   -subsystem_args 1
\end{lstlisting}

The last argument `subsystem\_args 1' signifies that only first order Legendre polynomials are used in constructing a total of $4 \times 3$ space-warping CG variables (sec. \ref{sub:Coarse-Graining-and-ss}). Once the hdf file 1PRT.hdf was created, a multiscale simulation was initiated with the following command:

\begin{lstlisting}[language=bash,tabsize=4, showstringspaces=false]
python -m proto_md run 1PRT.hdf
\end{lstlisting}

The simulation performs a series of $7,400$ CG steps, writing the minimized, equilibriated, and extrapolated coordinates of all protein atoms at the end of each CG timestep. The protein undergoes a conformational change over this time period, which is visualized by plotting the root-mean-square deviation (RMSD) of the protein atomic positions with the reference configuration set to the protein positions at $t=0$ ns. For comparison, two MD runs were also done (starting from the same initial conditions) at the same thermodynamic conditions.

\begin{lstlisting}[language=Python, showstringspaces=false]
import proto_md
import numpy as np
from MDAnalysis.analysis.align import *

sys = proto_md.System(`1PRT.hdf')

# create a new universe object, this is now
# set to the initial coordinates
mob = s.universe

# create a reference
ref = proto_md.System(`1PRT.hdf')

for ts in sys.timesteps:
	# align the mobile to the reference structure
	alignto(mob, ref, select="protein", mass_weighted=True)
	# save rmsd to file
	rmsd_list.append(rmsd(mob, ref))

# write rmsd to file
rmsd_list = np.array(rmsd_list)
np.savetxt('rmsd.dat', rmsd_list)
\end{lstlisting}

\begin{figure}[h]
\centering
\includegraphics[scale=0.6]{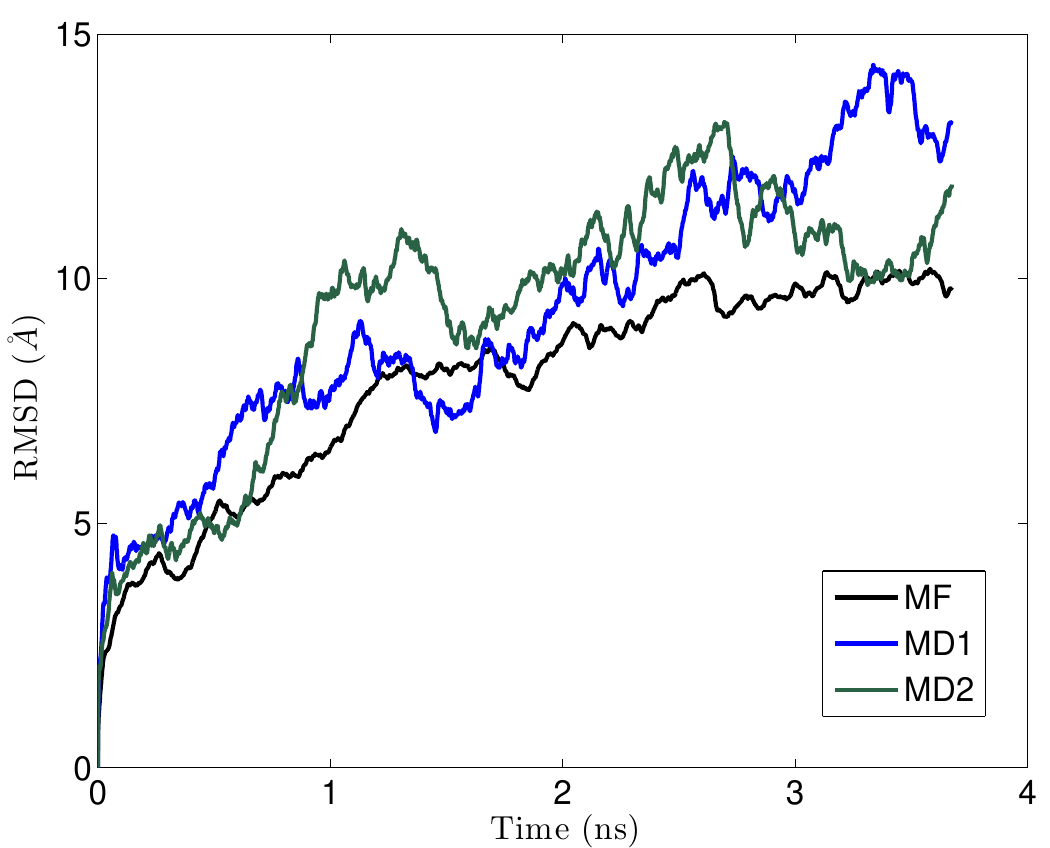}
\caption{Variation in the RMSD of pertussis toxin protein as a function of time, using multiscale factorization (MF) and molecular dynamics (MD1 and MD2).}
\label{fig:rmsd}
\end{figure}

%% file: Acknowledgments.tex
\section{Acknowledgments\label{sec:Acknowledgments}}
This research was supported in part by the following: the NSF INSPIRE program
(grant 1344263), the NSF division of material science research (grant 1533988), and the Indiana University information technology services (UITS) for high
performance computing resources.